\documentclass{PoS}

\title{An $ep$ collider based on proton-driven plasma wakefield acceleration}

\ShortTitle{Plasma wakefield collider}

\author{\speaker{Matthew WING}\thanks{Also affiliated with DESY, Hamburg.} \\
        UCL\\
        E-mail: \email{m.wing@ucl.ac.uk}
        }

\author{G. Xia, O. Mete \\
        University of Manchester, Cockcroft Institute
        }

\author{A. Aimidula, C. Welsch\\
        University of Liverpool and Cockcroft Institute
        }

\author{S. Chattopadhyay\\
        University of Lancaster, University of Liverpool, University of Manchester and Cockcroft Institute
        }
        
\author{S. Mandry\\
        Max Planck Institute for Physics, Munich and UCL
        }        

\abstract{
Recent simulations have shown that a high-energy proton bunch can excite strong 
plasma wakefields and accelerate a bunch of electrons to the energy frontier 
in a single stage of acceleration.  This scheme could lead to a future 
$ep$ collider using the LHC for the proton beam and a compact electron accelerator 
of length 170\,m, producing electrons of energy up to 100\,GeV.  The 
parameters of such a collider are discussed as well as conceptual layouts 
within the CERN accelerator complex.  The physics of plasma wakefield 
acceleration will also be introduced, with the AWAKE experiment, a proof 
of principle demonstration of proton-driven plasma wakefield acceleration, 
briefly reviewed, as well as the physics possibilities of such an $ep$ collider. 
}

\FullConference{XXII. International Workshop on Deep-Inelastic Scattering and Related Subjects,\\
		28 April - 2 May 2014\\
		Warsaw, Poland}

\begin{document}

\section{Introduction}

A high (TeV-scale) energy electron--proton collider would complement the proton--proton physics 
programme at the LHC and the planned electron--positron physics programme at the ILC.  The rich 
physics programme of the Large Hadron--Electron Collider (LHeC) is given in detail elsewhere~\cite{LHeC}, 
with only a brief discussion given here.  The cross section for Higgs production at LHeC is similar to that 
at the ILC and so with sufficiently high luminosity ($10^{34}$\,cm$^{-2}$\,s$^{-1}$), precise measurements 
of in particular the triple-gauge boson couplings can be made.  Measurements of inclusive deep inelastic 
scattering at these high scales should allow a full flavour decomposition of the parton densities in the 
proton and eliminate assumptions currently used as well as significantly reducing the uncertainty on 
their determination, often a limiting factor in the search for exotic physics at the LHC.  As well as considering 
the highest energy scales, the property of saturation at the very lowest Bjorken $x$ values will be probed.  
At low $x$, the nuclear structure is very poorly known and so an $eA$ physics programme will investigate 
an unmeasured region for $x < 0.01$.

In order to investigate such a wide-ranging physics programme of precision and potential discovery, the 
LHeC would use the 7\,TeV proton beam from the LHC and requires a new electron accelerator with 
energies of about 60\,GeV, giving a centre-of-mass energy of 1.3\,TeV.  To achieve such an electron energy, 
the favoured solution, based on conventional acceleration using radio frequency cavities, is a 9\,km long 
``racetrack" complex.  Such a large accelerator is needed due to the limitation in the maximum accelerating 
gradient to about 100\,MV\,m$^{-1}$ due to the onset of ionisation of the metal cavities.  Given this size, it is 
sensible to look at alternative technologies which could significantly reduce the length of the electron 
accelerator.  In these proceedings, such a technology, based on proton-driven plasma wakefield acceleration, 
is described.  Plasma wakefield acceleration is a scheme originally proposed at the end of 
the 1970s~\cite{prl:43:267} and is, in principle, applicable to all uses of accelerators and not just the LHeC.  
The use of proton-driven plasma wakefield acceleration has applications in general to the acceleration of particles to high 
energies~\cite{nim:a740:173} and so could also lead to a future $e^+e^-$ machine of much reduced size.

\section{Proton-driven plasma wakefield acceleration}

Plasma wakefields occur when a drive beam, a laser pulse or particle beam, enters a plasma and disturbs the free 
electrons.  In the case of a proton beam, the free electrons are attracted to the proton bunch, accelerate towards it, 
overshoot, are attracted back by the region of high positive charge density formed by the stationary ions, and hence 
form an oscillating system which creates large electric fields with an accelerating gradient in the direction of the 
incoming beam.  In the case of a laser pulse or an electron bunch as the driver, accelerating gradients of up to 
100\,GV\,m$^{-1}$ have been observed~\cite{pwa}.

Given the limitations of the initial energies of the laser pulse or electron beam, multiple acceleration stages would 
be required in order to accelerate electrons to the scales required for energy frontier machines.  As current proton 
beams have up to O(100)\,kJ of energy, the beam can propagate for long distances in a plasma and 
so act as a powerful driver of plasma wakefields.  Simulations demonstrated that electrons could be accelerated 
from 10\,GeV to 500\,GeV in about 300\,m of plasma using a proton beam of 1\,TeV~\cite{nphys:5:363}, with a 
maximum accelerating gradient of 3\,GV\,m$^{-1}$.

\section{The AWAKE experiment at CERN}

The Advanced Wakefield (AWAKE) collaboration was formed to perform a proof-of-principle experiment 
at CERN, showing that protons can drive strong plasma wakefields and that these can be used to accelerate 
electrons, in an initial phase, up to the GeV-scale in under 10\,m~\cite{AWAKE}.  The AWAKE experiment 
will use the SPS beam with an energy of 400\,GeV to drive the wakefields, however, in strong contrast to the 
concept of proton-driven plasma wakefield acceleration~\cite{nphys:5:363}, where the beam length considered 
was $\sigma_z = 100$\,$\mu$m, the length of the SPS proton bunch is 12\,cm.  Given the strength of the 
wakefield generated is proportional to $1/\sigma_z^2$, the strength of the wakefield for the SPS beam is 
potentially very low.  This is overcome by relying on the self-modulation instability (SMI)~\cite{prl:104:255003} in 
which transverse fields in the plasma split the long proton bunch into micro-bunches, spaced at the plasma 
wavelength.  These higher densities act constructively to create wakefields with an accelerating gradient of 
about 1\,GV/m. Witness electrons will be injected behind the proton bunch and a fraction of the electrons (up 
to 10\% of the bunch) will be accelerated from about 16\,MeV to about 2\,GeV in 6\,m of plasma.

\begin{figure}[hb]
\includegraphics[width=\textwidth]{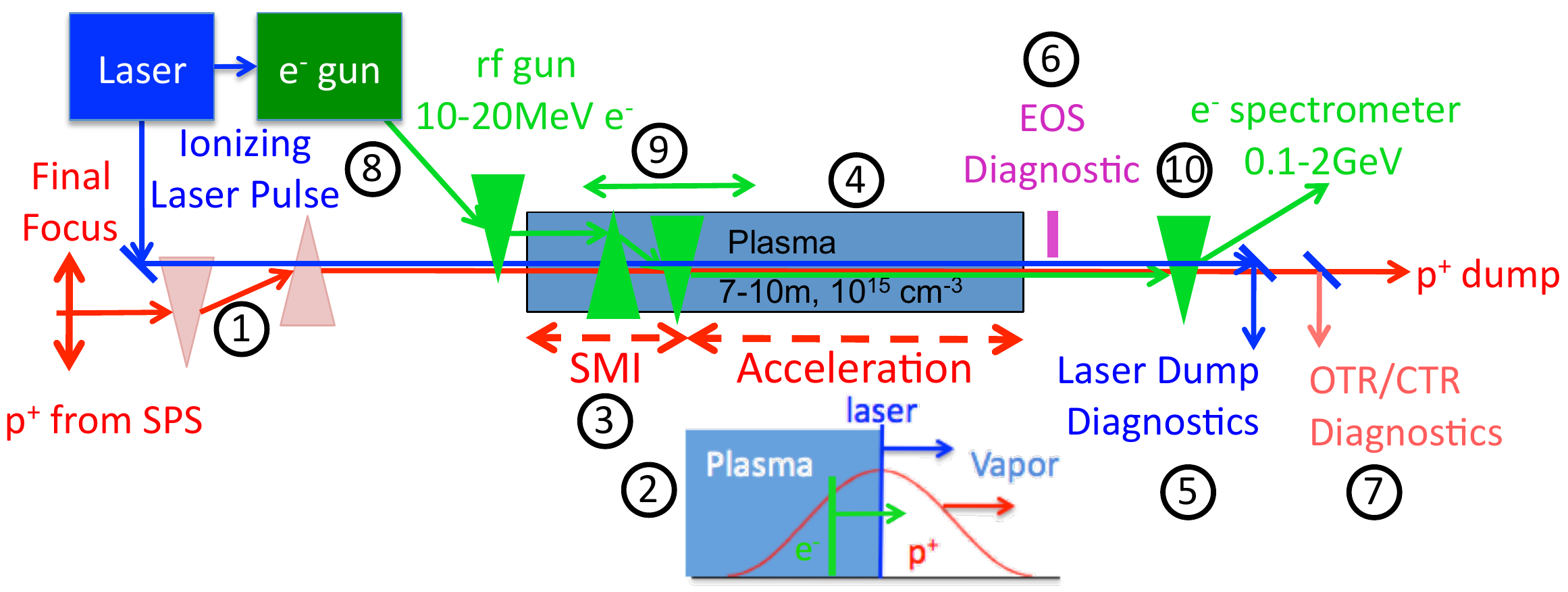} 
 \caption{Baseline design of the AWAKE experiment: The proton bunch extracted from the SPS is injected with 
the ionising laser pulse (1).   The laser pulse and the proton bunch travel together through the metal vapor cell.   
The co-propagating ionisation front shown in (2) provides the seeding of the SMI.   Growth of the SMI and the 
resulting self-modulation of the proton bunch occurs over the first $3-5$\,m of plasma (3).   The modulated bunch 
resonantly drives wakefields over the remaining length of plasma (4).   The laser pulse is dumped (5) and the 
proton bunch radial modulation is measured using electro-optical sampling (EOS) diagnostics (6) and optical 
transition radiation diagnostics (7).   An RF-gun driven by a laser pulse derived from the ionizing laser produces 
a witness electron bunch (8).   The electron bunch can be side-injected into the wakefield after the SMI has 
saturated (9).  Downstream of the plasma, the electron bunch energy spectrum is measured using a broad 
acceptance magnetic spectrometer (10). 
}
\label{fig:awake}
\end{figure}
The AWAKE experiment will be housed in the CNGS facility at CERN.  The general layout of the experiment 
is shown in Fig.~\ref{fig:awake}.   The proton beam propagates through a 10\,m long plasma cell, excites the 
wakefield and becomes modulated by this wakefield.  The short laser pulse propagates collinearly with the 
proton beam and serves the dual function of creating the plasma and seeding the SMI.  The electron bunch 
enters the plasma cell parallel to the proton beam with an initial offset of about 1\,cm and is merged into the 
wakefield several metres downstream as soon as the proton beam is modulated by the SMI.  A configuration 
in which the electron beam is collinear with the proton beam is now considered the default mode of running 
and although a beam of lower energy and larger energy spread is expected, this mode will have a higher 
capture efficiency.  Modulation of 
the proton beam radius is measured by electro-optical sampling (EOS) and optical and coherent transition 
radiation (OTR/CTR) diagnostics.   The accelerated electron beam is characterised with a magnetic 
spectrometer.

\section{An $ep$ collider: design and issues}

\begin{figure}
\includegraphics[width=\textwidth]{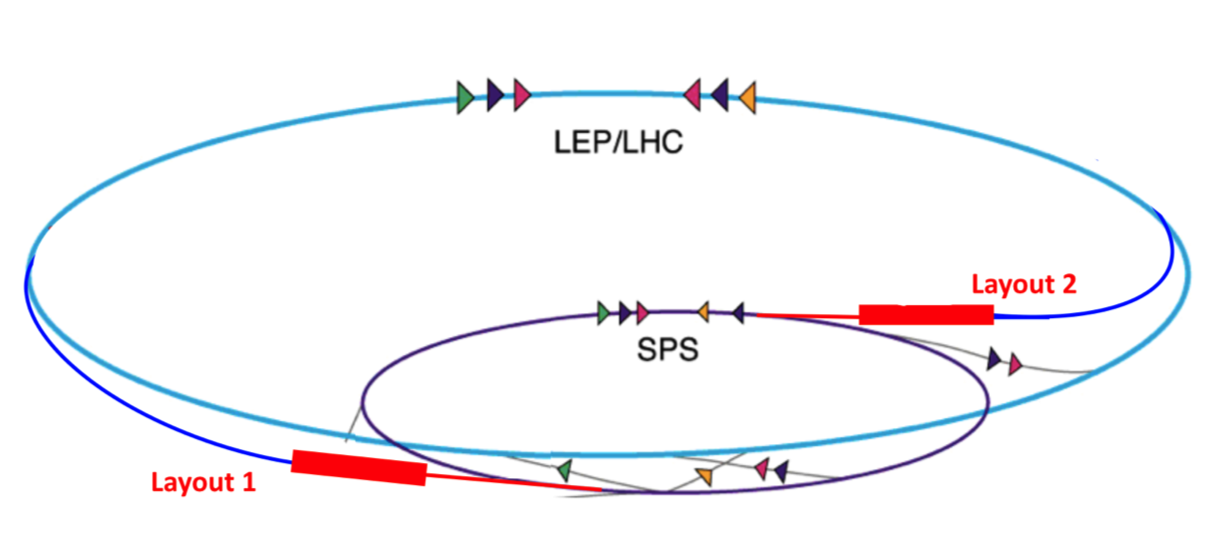} 
 \caption{A schematic of a  possible $ep$ collider using the CERN accelerator complex and the SPS proton 
 beam to drive plasma wakefield acceleration.  Electrons would be accelerated in the wakefields and collide with 
 protons in the LHC.}
\label{fig:ep}
\end{figure}

Assuming that the AWAKE experiment will demonstrate proton-driven plasma wakefield acceleration and that 
simulations correctly describe the physics of the process, application of the accelerator technology can be 
considered for a future $ep$ collider.  Aiming for similar energies to those proposed for the LHeC, a schematic 
of a possible $ep$ collider is shown in Fig.~\ref{fig:ep}.  The SPS proton beam is used to drive the wakefield and 
accelerate a trailing electron bunch to 100\,GeV in 170\,m of plasma.  These electrons then collide with the LHC 
proton beam, creating an $ep$ mode which runs parasitically with the LHC proton--proton collisions.  The basic 
sub-systems of the $ep$ collider are: the transfer and matching of protons to the plasma section; the electron 
source; the plasma section; the beam delivery and final focus; and the beam dump and/or recycling.  As this 
scheme utilises existing CERN infrastructure, there is the prospect of realising a high energy, cost-effective 
collider.

Although on the face of it this concept is very attractive, with a centre-of-mass energy of $\sqrt{s} = 1.67$\,TeV 
achieved with the a relatively short new electron beam line, there are a number of issues and challenges to 
be overcome in order to make this realisable~\cite{nim:a740:173}.

Phase slippage (or dephasing) occurs because there are particles travelling at different velocities and is a limiting 
factor of proton-driven plasma wakefield acceleration.  For a velocity of the wakefield which is the velocity of the 
proton driver, $\gamma_p$ and a velocity of the accelerating electrons, $\gamma_e$, the value of $\gamma_e$ can 
soon be greater than $\gamma_p$ and so the electrons overrun the wakefield and hence dephase.  The distance 
of the plasma accelerating section is therefore limited and for the SPS beam this is 170\,m; for the LHC beam as the 
driver, due to the higher energy, it is about 4\,km with electrons accelerated to 1\,TeV.  Such a scheme where the LHC 
is the drive beam can be used for an $e^+e^-$ linear collider with collisions at the TeV energy scale.

The proton beam will at some point become too spread and will not be able to drive a strong wakefield.  A way to 
compensate for this is to use external quadrupole magnets which will provide transverse focusing of the beam.  Additionally, 
the wakefields themselves have a focusing component and this may be enough to guide the proton beam.  Also given 
the highly relativistic nature of the proton beams, variations of the momentum spread will not be significant over the 
lengths being considered.  

The witness electrons can scatter off the plasma ions or electrons.  To assess this effect in detail, a model is being 
developed using a plasma simulation code and GEANT.

At AWAKE, the witness bunch will be electrons, but a similar controlled acceleration of positrons is necessary for a 
future linear collider, but also preferable at an $ep$ collider where the possibility to change between an electron and 
positron beam allows the electroweak sector to be probed.  Recent simulations have shown that a bunch of positrons  
can also be continuously accelerated to the TeV scale in a $\sim$ 1\,km long plasma~\cite{scirep:4:4171}.

The luminosity of an $ep$ machine is given by

\[
\mathcal{L}_{ep} = \frac{P_e \, N_p \, \gamma_p}{4\pi \, E_e \, \epsilon_p^N \, \beta_p^*} \,,
\]
where $E_e$ is the electron beam energy, $N_p$ is the number of particles in the 
proton bunch, $\epsilon_p^N$ is the normalised emittance of the proton beam, $\gamma_p$ is the Lorentz factor and 
$\beta_p^*$ is the beta function of the proton beam at the interaction point.  The electron beam power, $P_e$, is given 
by 

\[
P_e = N_e \, E_e \, n_b \, f_{\rm rep} \,,
\]
where $N_e$ is the number of particles in the electron bunch, $n_b$ is the number of bunches in the linac pulse 
and $f_{\rm rep}$ is the repetition rate of the linac.  Using the LHC beam parameters, \mbox{$N_p = 1.15 \times 10^{11}$}, 
$\gamma_p = 7460$, $\beta_p^* = 0.1$\,m, $\epsilon_p^N = 3.5\,\mu$m, and assuming electron beam parameters, 
$N_e = 1.15 \times 10^{10}$, $E_e = 100$\,GeV, $n_b = 288$ and $f_{\rm rep} = 15$, gives a luminosity of, 
\mbox{$\mathcal{L}_{ep} = 1 \times 10^{30}$\,cm$^{-2}$ s$^{-1}$}.  This luminosity value is significantly below that of conventional 
LHeC designs and so raises the question as to whether this can be increased, by e.g. increasing the repetition rate 
or decreasing the size of the electron beam at the interaction point.  Alternatively, physics at high energy, but lower luminosity should be 
considered.  As plasma wakefield acceleration has clearly demonstrated high accelerating gradients, there are prospects 
for a future $ep$ of $e^+e^-$ collider at the high energy frontier, but possibly with reduced luminosity than can be achieved 
with RF acceleration.  Brief studies have started~\cite{bartels-et-al} considering the physics that could 
be investigated at such colliders, such as classicalisation in electroweak and gravity, and should be further pursued.

\section{Summary}

The concept of proton-driven plasma wakefield acceleration and its application to a future $ep$ facility has been 
presented.  A proof-of-principle experiment, AWAKE, will start taking data in 2016 to demonstrate proton-driven 
plasma wakefield acceleration for the first time.  Based on such a scheme, simulations show that the current 
CERN accelerator complex could be used to generate a 100\,GeV electron beam in about 170\,m and along with 
the LHC proton beam have $ep$ collisions at a centre-of-mass energy of 1.67\,TeV.  Many challenges remain before 
such a collider could be realised, such as high luminosity and the acceleration of positrons, but further studies and 
the results of the AWAKE experiment will help to address these challenges.

\end{document}